\documentclass[11pt,a4paper,onecolumn,reqno]{amsart}
\usepackage[a4paper, margin=2.3cm, bmargin=3cm]{geometry}

\usepackage{graphicx,stfloats} \usepackage{color}
\usepackage{amsmath}
\usepackage{amsaddr}
\usepackage{bm, soul}
\usepackage{mathtools}
\usepackage[colorinlistoftodos,prependcaption,textsize=tiny]{todonotes}
\usepackage{braket}
\usepackage{hyperref}
\usepackage[square,comma,sort&compress, numbers]{natbib}
\usepackage{enumerate}
\usepackage[version=4]{mhchem}

\usepackage{placeins}

\renewcommand{\st}[1]{}

\usepackage{amssymb}
\usepackage{siunitx}
 
\usepackage{multirow}

\usepackage{etoolbox}
\patchcmd{\pprintMaketitle}
  {\hrule\vskip12pt}
 {\hrule\vskip12pt\ifvoid\extrainfobox\else\unvbox\extrainfobox\par\vskip12pt\fi}
 {}{}

\newsavebox\extrainfobox

\usepackage{caption}
\title{Assessment of flamelet manifolds for turbulent flame-wall interactions in Large-Eddy Simulations}

\author[stfs]{Yujuan Luo$^{1,*}$, Matthias Steinhausen$^{1}$, Driss Kaddar$^{1}$, Christian Hasse$^{1}$, Federica Ferraro$^{1}$}
\email{luo@stfs.tu-darmstadt.de} 
\address[]{$^1$Technical University of Darmstadt, Department of Mechanical Engineering, Simulation of reactive Thermo-Fluid Systems, Otto-Berndt-Stra{\ss}e 2, 64287 Darmstadt, Germany
}

\begin{document}
\pagestyle{plain}

\maketitle

\begin{abstract} 
A turbulent side-wall quenching (SWQ) flame in a fully developed channel flow is studied using Large-Eddy Simulation (LES) with a tabulated chemistry approach. Three different flamelet manifolds with increasing levels of complexity are applied: the Flamelet-Generated Manifold (FGM) considering varying enthalpy levels, the Quenching Flamelet-Generated Manifold (QFM), and the recently proposed Quenching Flamelet-Generated Manifold with Exhaust Gas Recirculation (QFM-EGR), with the purpose being to assess their capability to predict turbulent flame-wall interactions (FWIs), which are highly relevant to numerical simulations of real devices such as gas turbines and internal combustion engines. 

The accuracy of the three manifolds is evaluated and compared a posteriori, using the data from a previously published flame-resolved simulation with detailed chemistry for reference. For LES with the FGM, the main characteristics such as the mean flow field, temperature, and major species can be captured well, while notable deviations from the reference results are observed for the near-wall region, especially for pollutant species such as \ce{CO}. In accordance with the findings from laminar FWI, improvement is also observed in the simulation with QFM under turbulent flow conditions.
Although LES with the QFM-EGR shows a similar performance in the prediction of mean quantities as LES with QFM, it presents significantly better agreement with the reference data regarding instantaneous thermo-chemical states near the quenching point. This indicates the necessity to take into account the mixing effects in the flamelet manifold to correctly capture the flame-vortex interaction near the flame tip in turbulent configurations. Based on the findings from this study, suitable flamelet manifolds can be chosen depending on the aspects of interest in practical applications.
\end{abstract}


\section{Introduction} \addvspace{10pt}
\label{sec:introduction}
Combustion is a major way to generate power in modern society. Practical applications are usually operated in turbulent conditions. This results in a multi-physics phenomenon covering a wide range of length and time scales. When the combustion occurs in a confined space, the process becomes even more complicated due to flame-wall interactions (FWIs)~\cite{dreizler_advanced_2015}. Strong heat losses occur at the combustor wall, which can alter the flame structure and lead to flame quenching. This results in a lowered combustion efficiency and increased pollutant formation.

Turbulent FWI has a practical relevance for internal combustion devices such as power plant burners, gas turbines, and internal combustion engines. Therefore, extensive numerical investigations have been conducted to understand and model the underlying processes based on generic configurations, including both head-on quenching (HOQ)~\cite{pantangi_les_2014, lai_effects_2016, lai_turbulent_2017, lai_flow_2018,zhao_analysis_2018, zhao_effects_2021, LAI2022108896} and side-wall quenching (SWQ)~\cite{Alshaalan_2002,gruber_turbulent_2010,heinrich_large_2018,heinrich_investigation_2018, steinhausen_turbulent_2022, JIANG2021111432, ahmed2021scalar, ahmed2021influence} flames.
Among them, studies using direct numerical simulations (DNS) can be found in~\cite{Alshaalan_2002, gruber_turbulent_2010, lai_effects_2016, lai_turbulent_2017, lai_flow_2018, zhao_analysis_2018, zhao_effects_2021, LAI2022108896, steinhausen_turbulent_2022, JIANG2021111432, ahmed2021scalar, ahmed2021influence}. However, it must be noted that the current application of DNS mainly focuses on simple fuels such as \ce{H2}~\cite{gruber_turbulent_2010}, or simplified chemical mechanisms such as single-step Arrhenius type reaction mechanisms~\cite{Alshaalan_2002, lai_effects_2016, lai_turbulent_2017, lai_flow_2018, zhao_analysis_2018, zhao_effects_2021}, or simple geometries~\cite{LAI2022108896, steinhausen_turbulent_2022, JIANG2021111432, ahmed2021scalar, ahmed2021influence, kaddar2022}. Their application in real devices is usually prohibited due to the large computational cost. In this context, models based on chemistry manifolds can be a good alternative since they help to reduce the number of transport equations to be solved while maintaining a detailed chemistry representation. Such methods include the Flamelet/Progress Variable (FPV) model~\cite{coelho2001unsteady}, Flamelet-Generated Manifolds (FGMs)~\cite{oijen2000modelling}, Intrinsic Low-Dimensional Manifolds (ILDMs)~\cite{maas1992simplifying}, flame prolongation of ILDM (FPI)~\cite{gicquel2000liminar}, and Reaction-Diffusion Manifolds (REDIMs)~\cite{bykov2007extension}. With these methods, thermo-chemical states are pre-calculated and stored in look-up tables that are accessed by the control variables during the CFD simulation.
Since FWI presents unique characteristics compared to unconfined combustion, these models need to be extended and validated before they can be applied to the simulation of near-wall combustion processes in real devices. For example, complex physics such as the substantial heat losses and the transient effects in turbulent flows needs to be additionally accounted for.

The earliest attempts to describe the heat loss in FWI employed FGM, with the lower limit of the enthalpy level extended by burner stablized flamelets. Heinrich et al.~\cite{heinrich_3d_2018} performed both two-dimensional and three-dimensional simulations of the SWQ of a laminar premixed \ce{CH4}-air flame, which was experimentally measured by Jainski et al.~\cite{jainski2017sidewall}.
The results show that FGM can reproduce the flame structure correctly and capture major characteristics such as \ce{CO2} and temperature. However, the near-wall \ce{CO} is significantly underpredicted. The unsuitability of FGM in \ce{CO} prediction was also found by Ganter et al.~\cite{ganter_numerical_2017}, who performed simulations on a two-dimensional subdomain of the configuration in \cite{heinrich_3d_2018} with a finer grid resolution. They also concluded that the near-wall \ce{CO} accumulation results from transport originating from \ce{CO} produced in regions away from the wall. 
To explore the root cause of the deficiencies of FGM, SWQ processes were taken into account in the composition space in~\cite{ganter_laminar_2018} to quantify the deviations from exact solutions employing scalar dissipation rates. 
It was found that the discrepancies in \ce{CO} predicted by FGM are due to improper considerations of species diffusion in the enthalpy direction. 
To remedy this drawback, an improved chemistry manifold was developed: a REDIM based on one-dimensional detailed chemistry head-on quenching (HOQ) solutions. Similarly, Efimov et al.~\cite{efimov_qfm_2020} proposed a Quenching Flamelet-Generated Manifold (QFM) to improve the accuracy of \ce{CO} prediction in FWI. Based on HOQ flamelets, the effects of scalar dissipation in the direction of enthalpy can be included in the manifold. The method was validated against a detailed chemistry simulation of the SWQ of a two-dimensional premixed laminar \ce{CH4}-air flame. A comparison of the performance of REDIM and QFM was conducted by Steinhausen et al.~\cite{steinhausen2021numerical}, based on the SWQ of laminar \ce{CH4}-air and DME-air flames that were experimentally investigated by Kosaka et al.~\cite{kosaka_wall_2018,kosaka_effect_2020}. It was observed that REDIM and QFM do not exhibit a significant difference in \ce{CO} prediction, and both are more accurate than FGM. Luo et al.~\cite{luo_simulation_2021,luo2022manifold,luo2022model} proposed the REDIM method formulated and constructed in generalized coordinates for FWI processes. The method performs well in predicting the SWQ of the laminar {\ce{CH4}}-air flame experimentally studied in~\cite{jainski2017sidewall}, and the SWQ of the laminar DME-air flame measured by Zentgraf et al.~\cite{zentgraf2022detailed}. Note that the advantages of REDIM and QFM compared with FGM have mostly been proven in laminar SWQ flames. However, to the best of the authors' knowledge, these improved manifolds have not yet been validated and assessed in coupled simulations of turbulent FWI. For the Large-Eddy Simulation (LES) of turbulent FWI, FGM is the only method reported in the literature. For example, Heinrich et al.~\cite{heinrich_large_2018,heinrich_investigation_2018} performed LESs on an experimental SWQ burner~\cite{jainski_experimental_2017} using FGM, coupled with the artificial thickened flame (ATF) approach~\cite{colin2000thickened,kuenne2011modeling}. Donini et al.~\cite{donini_numerical_2013}, Proch et al.~\cite{Proch_modeling_2015}, and Tang et al.~\cite{tang_large_2021} simulated a generic lab scale burner~\cite{lammel_experimental_2012} for high-velocity preheated jets with FGM considering heat losses. These studies only investigated velocities and global quenching quantities; information related to pollutant formation was not reported. 
Pantangi et al.~\cite{pantangi_les_2014} conducted simulations on an experimental HOQ burner \cite{mann_transient_2014} using similar methods, and found that the near-wall \ce{CO} prediction shows discrepancies from the experimental data. 
This is consistent with the findings from the laminar studies~\cite{ganter_numerical_2017, efimov_qfm_2020,steinhausen2021numerical}.

In turbulent FWI, besides heat loss, the effects of turbulent mixing have to be considered. Jiang et al.~\cite{JIANG2021111432} performed a three-dimensional DNS of FWI in the case of \ce{CH4}-air flames diluted by hot burned products. It was shown that solutions of one-dimensional freely propagating flames cannot describe the variation of \ce{CO} in the region close to the wall. This was attributed to turbulent mixing and diffusion effects.
The effect of turbulent mixing was also reported in a recent experimental study conducted by Zentgraf et al.~\cite{zentgraf_classification_2021}. Similarly to the findings in ~\cite{palulli2019unsteady,heinrich_investigation_2018}, both SWQ-like and HOQ-like behaviors were observed. Focusing on the SWQ-like scenarios, thermo-chemical states that have not been encountered in laminar conditions were observed in the near-wall region. To explain the phenomenon, a flame-vortex interaction (FVI) mechanism was proposed. According to this explanation, the flame interacts with the near-wall vortical structures, entraining the burned gases into the fresh gas mixture near the flame’s quenching point.
More recently, this mechanism was confirmed in a numerical study based on a three-dimensional flame-resolved simulation  (FRS) with detailed chemistry of the turbulent FWI of a stoichiometric \ce{CH4}-air flame in a fully developed channel flow~\cite{steinhausen2022flame}. To take into account the effects of flame dilution due to exhaust gas recirculation in the FVI area, as well as substantial heat losses to the wall, a novel flamelet manifold, the QFM with Exhaust Gas Recirculation (QFM-EGR) was proposed. Moreover, the QFM and QFM-EGR were validated in an \textit{a-priori} manner.

Although there are plenty of studies focusing on the validation of manifold-based models in turbulent combustion~\cite{Nguyen_multidimensional_2010,popp2015flamelet,vanoijen_state_2016}, the investigation of chemistry manifolds for turbulent FWI is rather rare. It has been shown for laminar FWI that the choice and the suitability of the manifold are crucial to capturing physical processes such as pollutant formation near the wall in fully coupled simulations, however this has not been investigated in turbulent FWI yet. To fill this gap, the performance of chemistry manifolds for FWI: FGM, QFM~\cite{efimov_qfm_2020}, and QFM-EGR~\cite{steinhausen2022flame} are evaluated in the LES of a turbulent \ce{CH4}-air SWQ, using the FRS performed by Steinhausen et al.~\cite{steinhausen_turbulent_2022} as a reference. The two main objectives of the paper are:
\begin{enumerate}
    \item to assess the predictive capabilities of the existing flamelet manifolds for FWI in the context of LES. To our knowledge, this is the first attempt to apply both the QFM and QFM-EGR in a coupled LES, and compare the performance of FGM, QFM, and QFM-EGR following an \textit{a-posteriori} approach;
    \item to evaluate the relevance and statistical importance of the FVI mechanism in turbulent FWI, based on comparison between QFM and QFM-EGR, with the latter taking into account the effects of burned gas recirculation.
\end{enumerate}

In the following, the numerical methods adopted are introduced in Section~\ref{sec:NumericalMethods}, including descriptions of the FGM, QFM, QFM-EGR, and the governing equations for the coupled simulations. The numerical configuration is outlined in Section~\ref{sec:NumericalSetup}. Section~\ref{sec:ResultsAndDiscussion} contains results and discussions. Firstly, LES results for FGM, QFM, and QFM-EGR are presented and compared with data from the FRS. Afterwards, the influences of the FVI mechanism are discussed. Finally, conclusions are drawn in Section~\ref{sec:Conclusions}.

\section{Numerical methods} \addvspace{10pt}
\label{sec:NumericalMethods}
In this study, a turbulent, premixed \ce{CH4}-air flame ignited in a fully developed channel flow undergoing SWQ at the channel wall is simulated using LES. The computations are performed with a reduced version of the CRECK mechanism \cite{RANZI2012468}, which was reduced ad hoc for FWI conditions using the approach described in \cite{Stagni2016} and consists of 24 species and 165 chemical reactions. The mechanism was designed to capture all relevant combustion physics in the investigated configuration. The unity Lewis number assumption is adopted for the calculation of molecular diffusion. To enable comparison between different models, the choice of the chemical mechanism and the transport model is consistent with the FRS studies~\cite{steinhausen_turbulent_2022, steinhausen2022flame}. Note that the chemistry manifold is not restricted to the choice of the reaction mechanism.

Turbulent combustion is modeled with flamelet manifolds in combination with the ATF approach. Specifically, three flamelet manifolds are assessed: (1) FGM~\cite{oijen2000modelling, ketelheun2013heat}, (2) QFM~\cite{efimov_qfm_2020}, and (3) QFM-EGR~\cite{steinhausen2022flame}, an extension of QFM. The table generation procedure of the FGM and QFM follows the work of Steinhausen et al.~\cite{steinhausen2021numerical}, while the QFM-EGR table is generated similarly to~\cite{steinhausen2022flame}, as shown in Table~\ref{table3}. Note that the \textit{a-priori} results based on QFM-EGR already showed a very good agreement with the FRS~\cite{steinhausen2022flame}, therefore additional effects such as the flame stretch are not included in the manifolds. More details will be discussed in the following.

\begin{table}[htbp] 
\small
\centering
\caption{Characteristics of FGM, QFM, and QFM-EGR.}
\begin{tabular}{|>{\centering}p{2cm}| >{\centering}p{9.5cm}|>{\centering\arraybackslash}p{3.5cm}|}
\hline
 Tabulation strategy & Flamelet types  & Physics included \\
\hline
FGM & (1) 1D freely propagating flames with $T_\mathrm{inflow} \geqslant $ 300~K, \\(2) 1D freely propagating flames with different EGR levels and $T_\mathrm{inflow} =$ 300 K & Mixing-induced enthalpy variations\\
\hline
QFM & (1) 1D freely propagating flames with $T_\mathrm{inflow} >$ 300 K, \\(2) 1D HOQ flame with $T_\mathrm{inflow} =$ 300~K  & Quenching-induced heat losses\\
\hline
QFM-EGR & (1) 1D HOQ of 1D freely propagating flames with $T_\mathrm{inflow} \geqslant $~300 K, \\(2) 1D HOQ of 1D freely propagating flames with different EGR levels and $T_\mathrm{inflow} =$ 300 K & Quenching-induced heat losses and mixing with cooled exhaust gas \\
\hline  
\end{tabular}
\label{table3}
\end{table}

\subsection{Generation of flamelet manifolds} \addvspace{5pt}
\label{sec:manifoldGeneration}

To generate the FGM, a series of independent one-dimensional freely propagating flames with varying enthalpy levels are computed. Enthalpy reduction is achieved by exhaust gas recirculation, i.e., different amounts of cooled burned gases are added to the inflow mixture of the flame. The amount of burned gases is given by EGR\,=\,$m_\mathrm{bg}/(m_\mathrm{fg}+m_\mathrm{bg})$, where $m_\mathrm{bg}$ denotes the mass flux of the cooled burned gas and $m_\mathrm{fg}$ the mass flux of the fresh gases. Additionally, the upper enthalpy limit is expanded using one-dimensional freely propagating flames with EGR\,=\,0 and increased inflow temperatures. Finally, the spanned manifold  $\psi=\psi\left(x, \text{EGR}\right)$ is parameterized by a progress variable ($Y_\mathrm{PV}$) and the enthalpy ($h$), with the former one characterizing the progress of the chemical reaction and the latter the heat loss to the wall. Here $x$ is the spatial coordinate. Therefore, the enthalpy variation is decoupled from the reaction progress variable variation for FGM. Note that the choice of enthalpy-reduction method has only small influences on the resulting chemistry manifold~\cite{Oijen_2000}, indicating that other methods, such as one-dimensional burner stablized flames~\cite{fiorina2003modelling}, can also be used here.
When generating the QFM, in order to take into account the substantial heat loss to the wall, a single transient HOQ flame is tabulated, resulting in a thermo-chemical state $\psi=\psi\left(x, t\right)$, where $t$ is the time. The wall and the fresh gas temperature are set to 300~K, while the fresh gas composition is set according to the inflow conditions of the reference simulation. Similarly to the FGM, this two-dimensional manifold is mapped to a progress variable-enthalpy state, i.e., $\psi=\psi\left(Y_\mathrm{PV}, h \right)$, with the upper enthalpy limit extended by one-dimensional freely propagating flames with increased inflow temperatures. However, in this case, the enthlapy dimension and the reaction progress dimension are closely related by the quenching process. For efficient access to the manifold, the control variables of both the FGM and the QFM are normalized, namely
\begin{equation}
    h^* = \frac{h - h_\mathrm{min}}{h_\mathrm{max} - h_\mathrm{min}} \ ,
\end{equation}
\begin{equation}
    Y_\mathrm{PV}^* = \frac{Y_\mathrm{PV} - Y_\mathrm{PV, \, min}\left(h^*\right)}{Y_\mathrm{PV, \, max}\left(h^*\right) - Y_\mathrm{PV, \, min}\left(h^*\right)} \ .
\end{equation}
The thermo-chemical states are stored in a table as $\psi = \psi \left(Y_\mathrm{PV}^*, h^* \right)$ with a resolution of 101 $\times$ 101 for $Y_\mathrm{PV}^* \times h^*$.

To generate the QFM-EGR, a series of transient, one-dimensional HOQ flames are calculated. The transient HOQ flames are initialized with one-dimensional freely propagating flames with different fresh gas compositions, leading to different thermo-chemical states. The initial freely propagating flames consist of (1) 61 flames with varying EGR levels from 0 to 0.3 and the same inflow temperature, with the fresh gas mixture gradually diluted with cold exhaust gases, similarly to FGM, and (2) 19 preheated one-dimensional flames with increased inflow temperatures. Note that the temperature of the cooled exhaust gas is set to be similar to the fresh gas temperature since it is reported in \cite{zentgraf_classification_2021} that the burned gas cools down significantly by heat transfer to the wall and mixes with fresh gas upstream the quenching point during its passage in the gap between the flame tip and the wall. The 80 individual HOQ simulations result in a three-dimensional manifold $\psi=\psi(x,t,\text{EGR})$, which is then mapped to $\psi=\psi \left(Y_\mathrm{PV}, Y_\mathrm{PV2}, h \right)$. Consistent with \cite{steinhausen2022flame}, \ce{CO} is chosen as the second progress variable, i.e., $Y_\mathrm{PV2}=Y_{\ce{CO}}$, to account for the amount of exhaust gases added to the flame. Note that the choice of the second progress variable is not restricted to the \ce{CO} value, and theoretically any other variables that fulfill the condition of unique parametrization of the EGR dimension can be used. Similarly to before, normalization is applied to all control variables, which reads
\begin{equation}
    Y_\mathrm{PV}^* = \frac{Y_\mathrm{PV} - Y_\mathrm{PV, \, min}}{Y_\mathrm{PV, \, max} - Y_\mathrm{PV, \, min}} \ ,
\end{equation}
\begin{equation}
    Y_\mathrm{PV2}^* = \frac{Y_\mathrm{PV2} - Y_\mathrm{PV2, \, min}\left(Y_\mathrm{PV}^*\right)}{Y_\mathrm{PV2, \, max}\left(Y_\mathrm{PV}^*\right) - Y_\mathrm{PV2, \, min}\left(Y_\mathrm{PV}^*\right)} \ ,
\end{equation}
\begin{equation}
    h^* = \frac{h - h_\mathrm{min}\left(Y_\mathrm{PV}^*, Y_\mathrm{PV2}^* \right)}{h_\mathrm{max}\left(Y_\mathrm{PV}^*, Y_\mathrm{PV2}^* \right) - h_\mathrm{min}\left(Y_\mathrm{PV}^*, Y_\mathrm{PV2}^* \right)} \ .
\end{equation}

In this case, the thermo-chemical states stored in the flamelet table can be expressed as $\psi = \psi \left(Y_\mathrm{PV}^*, Y_\mathrm{PV2}^*, h^* \right)$. The manifold dimensions are 150 $\times$ 201 $\times$ 101 for $Y_\mathrm{PV}^* \times Y_\mathrm{PV2}^* \times h^*$.

\subsection{Governing equations for coupled simulations} \addvspace{5pt}
\label{sec:coupledSimulations}

Within the LES framework, the conservation equations for mass and momentum read 

\begin{equation}
    \frac{\partial \overline{\rho}}{\partial t} + \frac{\partial \overline{\rho} \widetilde{u}_i}{\partial x_i} = 0 \ ,
\end{equation}

\begin{equation}
    \frac{\partial \overline{\rho} \widetilde{u}_i}{\partial t} + \frac{\partial \overline{\rho} \widetilde{u}_i \widetilde{u}_j}{\partial x_j} = \frac{\partial }{\partial x_j}\left(\overline{\tau}_{ij}-\overline{\rho}\widetilde{u_i^{''} u_j^{''}}\right) - \frac{\partial \overline{p}}{\partial x_i} \ ,
\end{equation}
where $\rho$ is the density, $u_i$ the $i$-th component of the velocity, $u_i^{''}$ the $i$-th subgrid component of the velocity, $p$ the pressure, and $\tau_{ij}$ the component of the shear stress tensor. The operators $\overline{\cdot}$ and $\widetilde{\cdot}$ represent the filtering and the Favre filtering, respectively. The subgrid stresses $\widetilde{u_i^{''} u_j^{''}}$ are closed using the $\sigma$ model \cite{Nicoud2011,hubert2011dynamic}.

The ATF approach~\cite{colin2000thickened,kuenne2011modeling} is adopted to correctly capture the turbulence-chemistry interaction. For the coupled simulations, the governing equations for the control variables read 

\begin{equation}
    \frac{\partial \overline{\rho} \widetilde{Y}_{\mathrm{PV}}}{\partial t} + \frac{\partial \overline{\rho} \widetilde{u}_i \widetilde{Y}_{\mathrm{PV}}}{\partial x_i} = \frac{\partial}{\partial x_i}\left(FE \overline{\rho} D \frac{\partial \widetilde{Y}_{\mathrm{PV}}}{\partial x_i}\right) + \frac{\partial}{\partial x_i} [(1-\Omega) \overline{\rho} D_t \frac{\partial \widetilde{Y}_{\mathrm{PV}}}{\partial x_i}] + \frac{E}{F} \overline{\dot{\omega}}_{\mathrm{PV}} \ ,
    \label{eq:PV1}
\end{equation}

\begin{equation}
    \frac{\partial \overline{\rho} \widetilde{h}}{\partial t} + \frac{\partial \overline{\rho} \widetilde{u}_i \widetilde{h}}{\partial x_i} = \frac{\partial}{\partial x_i}\left(FE \overline{\rho} D \frac{\partial \widetilde{h}}{\partial x_i}\right) + \frac{\partial}{\partial x_i}[(1-\Omega) \overline{\rho} D_t \frac{\partial \widetilde{h}}{\partial x_i}] \ ,
    \label{eq:h}
\end{equation}

\begin{equation}
    \frac{\partial \overline{\rho} \widetilde{Y}_{\mathrm{PV2}}}{\partial t} + \frac{\partial \overline{\rho} \widetilde{u}_i \widetilde{Y}_{\mathrm{PV2}}}{\partial x_i} = \frac{\partial}{\partial x_i}\left(FE \overline{\rho} D \frac{\partial \widetilde{Y}_{\mathrm{PV2}}}{\partial x_i}\right) + \frac{\partial}{\partial x_i} [(1-\Omega) \overline{\rho} D_t \frac{\partial \widetilde{Y}_{\mathrm{PV2}}}{\partial x_i}] + \frac{E}{F} \overline{\dot{\omega}}_{\mathrm{PV2}} \ .
    \label{eq:PV2}
\end{equation}
Here, $\Omega$ denotes the flame sensor, which is used to avoid non-physical thickening outside the flame. The flame sensor employed here is an adaptation of the one in~\cite{heinrich_large_2018}, using a second-order polynomial with the maximum located at the maximum reaction rate of the progress variable for each enthalpy level. In the non-reactive regions of the flame, the flame sensor is blended to zero to avoid thickening.
$F$ is the thickening factor, and it is dynamically evaluated based on $\Omega$ and the grid-dependent maximum thickening factor $F_{\mathrm{max}}$~\cite{pantangi_les_2014}. $E$ is the efficiency function, which is defined as the ratio between the total flame surface and its resolved part in the filtered volume. It is introduced to compensate for the flame surface loss due to the thickening, and the value is calculated using the analytical model developed by Charlette et al.~\cite{CHARLETTE2002159}. $D$ is the molecular diffusivity of the scalar. $D_t$ is the subgrid diffusivity calculated as $D_t$ = $\mu_t$ / $Sc_t$, where $\mu_t$ is the turbulent viscosity and the turbulent Schmidt number $Sc_t$ is chosen as 0.4. For the first progress variable ($Y_\mathrm{PV}=Y_{\ce{CO2}}$), the reaction rate $\dot{\omega}_{\mathrm{PV}}$ is directly obtained from the flamelet manifold. For the second progress variable ($Y_\mathrm{PV2}=Y_{\ce{CO}}$), the reaction rate is calculated based on the production rate $\dot{\omega}^{+}$ and the consumption rate $\dot{\omega}^{-}$ taken from the flamelet manifold, namely $\dot{\omega}_{\rm{PV2}} = \dot{\omega}_{\mathrm{CO}}^{+} + Y_{\mathrm{PV2}} ({\dot{\omega}_{\mathrm{CO}}^{-}}/{Y_{\mathrm{CO}}^{\mathrm{FLT}}})$, allowing \ce{CO} to evolve based on its own time scale~\cite{ihme_modeling_2008}. Note that for FGM and QFM, equations only need to be solved for the first progress variable (Eq.~(\ref{eq:PV1})) and the enthalpy (Eq.~(\ref{eq:h})). For QFM-EGR, all three control variables need to be transported, i.e., Eqs.~(\ref{eq:PV1})-(\ref{eq:PV2}) need to be solved.

\section{Numerical setup} \addvspace{10pt}
\label{sec:NumericalSetup}
\begin{figure}[h]
\centering
\includegraphics[trim = 70mm 64mm 100mm 97mm, clip, width=350pt]{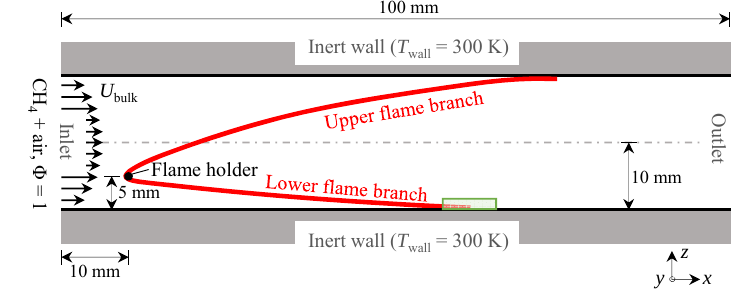}
\caption{Schematic of the configuration investigated. The~$x$, $y$, and $z$ coordinates are the streamwise, lateral, and wall-normal directions, respectively. The dimension in the lateral direction is 30 mm. The region of interest analyzed in Fig.~\ref{scatterTotal} is denoted by
the green rectangle.}
\label{configuration}
\end{figure}

The configuration has been studied by Steinhausen et al.~\cite{steinhausen2022flame, steinhausen_turbulent_2022} with an FRS. A V-shaped premixed flame stabilized by a flame holder in a fully developed turbulent channel undergoes FWI at the channel walls, as shown in Fig.~\ref{configuration}. The coordinates are defined as follows: $x$, $y$, and $z$ correspond to streamwise, lateral, and wall-normal directions, respectively. The inflow is a stoichiometric \ce{CH4}-air mixture at 300 K. The Reynolds number of the inert channel flow is $Re = (U_{\mathrm{bulk}}H)/\nu \approx 5540$, where $U_\mathrm{bulk}$ is the mean axial flow velocity (4.4 m/s), $H$ the channel height (20 mm), and $\nu$ the kinematic viscosity of the inflow.
Similarly to \cite{gruber_turbulent_2010}, the flame holder is modeled as a cylindrical region ($r$ = 0.9 mm) filled with burned gases. It is placed at a distance of $H$/4 from the bottom wall. Both the upper and lower walls are assumed to be inert and have a fixed temperature that is equal to the inflow temperature, namely $T_{\mathrm{wall}}$ = 300~K.

An inert turbulent channel flow is simulated in a pre-processing step to provide proper inflow boundary conditions for the reactive case. For the inert case, the dimensions of the computational domain are $7H$, $1.5H$, and $H$ in $x$, $y$, and $z$ directions, respectively. A stretched structured grid is adopted, which is uniform in $x$ and $y$ directions and refined towards the wall in $z$ direction. In the core flow, the non-dimensional grid resolution in $x$, $y$, and $z$ directions is $\Delta x^+ = \Delta y^+ = \Delta z^+$ = 5.14, where the superscript $^+$ denotes non-dimensionalization with the viscous length scale. The finest grid spacing in the $z$ direction is $\Delta z^+$ = 1. The total number of cells is 7.4 million. A no-slip boundary condition is applied to the channel walls, and a periodic boundary condition is used for lateral and streamwise boundaries. The inflow velocity field is used as the inflow boundary condition for the reactive case, and it is stored with a time interval of 17 $\mu$s.

Regarding the reactive case, the computational domain is similar to the inert case, except that the length is reduced from $7H$ to $5H$. With the same grid resolution as the inert case, the resulting total number of cells is 5.3 million. The boundary conditions employed are as follows: at the inlet, the inflow velocity field obtained from the inert case is spatially and temporally interpolated at every time step, which is about $\Delta t \approx$ 2.5 $\mu$s. At the outlet, a zero gradient boundary condition is applied for the velocity, progress variables, and enthalpy. At the channel walls, a no-slip boundary condition is employed for the velocity, a zero-gradient boundary condition for the progress variables, and a fixed value of 300 K for the temperature. In the lateral direction, periodic boundary conditions are adopted.

The numerical simulations are performed using an in-house solver~\cite{Weise2013} based on OpenFOAM~\cite{wellerTensorial1998} with second-order discretizations in time and space. 

\section{Results and discussion} \addvspace{10pt}
\label{sec:ResultsAndDiscussion}
In the following, the performances of FGM, QFM, and QFM-EGR are assessed by comparison with the FRS results from Steinhausen et al.~\cite{steinhausen2022flame, steinhausen_turbulent_2022}. Firstly, profiles of mean flow field quantities are analyzed. Afterwards, instantaneous results are presented to further show the capability of the models. Specifically, probability density functions (PDFs) of \ce{CO} are evaluated first. Then, the importance of FVI is studied based on cross-comparison of the simulation results.

Since the thermo-chemical states within the quenching region strongly depend on the streamwise position, a flame-fixed coordinate system is introduced, following previous studies~\cite{heinrich_3d_2018,ganter_numerical_2017,steinhausen2021numerical}.
The coordinates of the quenching point are defined as ($x_Q$, $y_Q$, $z_Q$). Similarly to previous works~\cite{gruber_turbulent_2010,heinrich_3d_2018}, the quenching point for each wall-normal plane is determined based on the maximum wall heat flux, which is evaluated at a near-wall horizontal plane $z$~=~50~$\mu m$ for consistency. The instantaneous wall heat flux is calculated according to 
\begin{equation}
    \lambda \left. \frac{\partial T}{\partial z} \right|_{z=50\mu m} \ ,
\end{equation}
where $\lambda$ corresponds to the gas phase thermal conductivity at $z~=~50~\mu m$. With the variation of the wall heat flux, the streamwise position of the quenching point $x_Q$ can be found. To further determine its lateral and wall-normal coordinates, position of the flame front is used. Similarly to~\cite{mann_transient_2014}, the flame front is defined based on an isothermal contour, i.e., $T$ = 1500~K.
Based on the quenching point, a relative coordinate system is defined. Within these coordinates, $y$ and $z$ remain unchanged and $x_q$ denotes the coordinate parallel with the wall determined by $x_q = x - x_Q$.

\subsection{Reactive scalars and mean flow field}  \addvspace{5pt}
\label{sec:globalFlameCharacteristics}

The major flame characteristics are assessed based on several mean flame quantities, which are obtained based on both time and space averaging of the instantaneous fields. Firstly, the instantaneous results are averaged over a sufficiently long time, i.e., more than 10 flow-through times, to ensure time independence. Afterwards, the time-averaged quantities are further averaged along the $y$ direction. These are referred to as mean quantities in the following, as denoted by $<\cdot>$.

\begin{figure}[h!]
\centering
\includegraphics[trim = 0mm 0mm 0mm 0mm, clip, width=350pt]{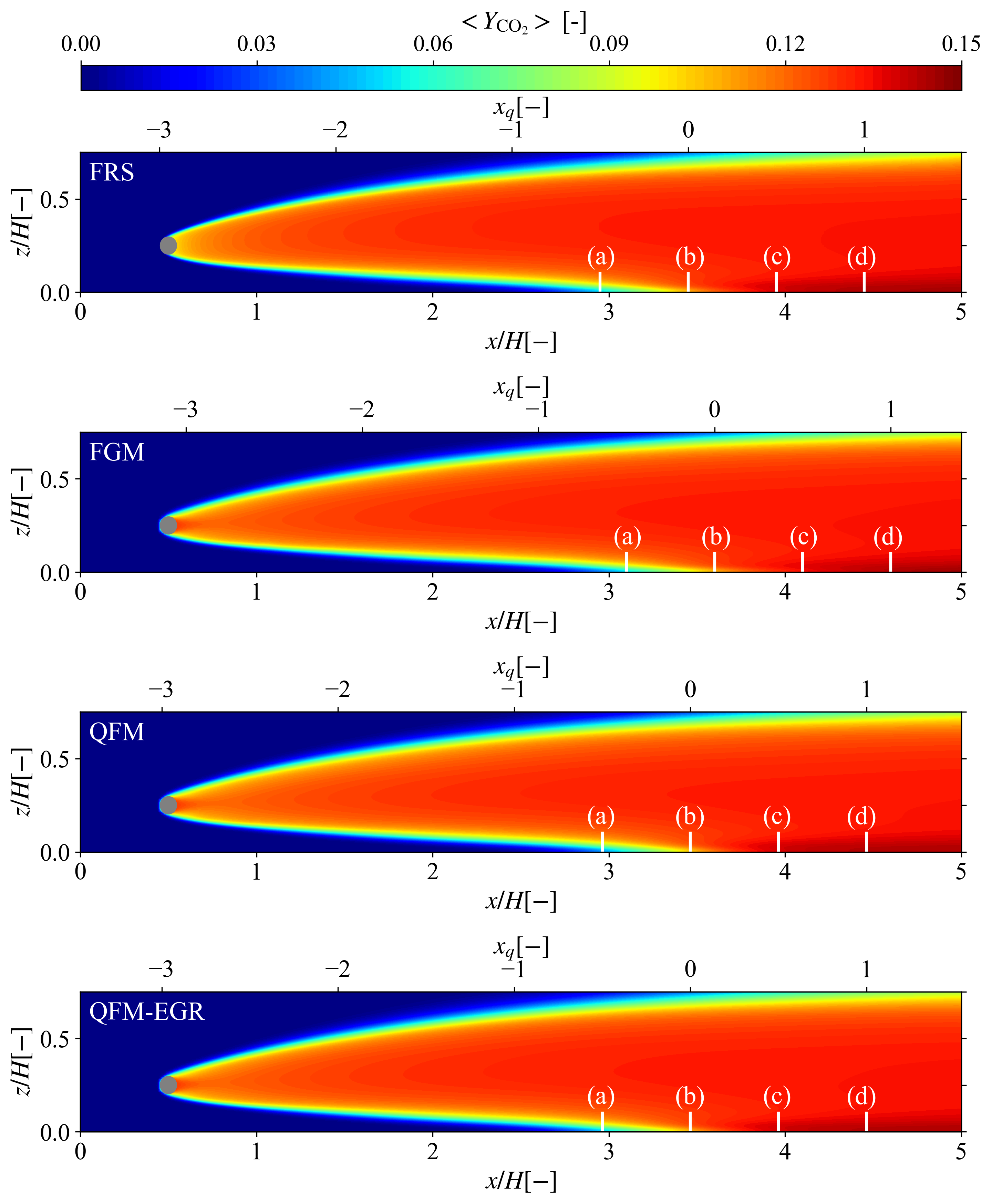}
\caption{Contour of mean $Y_{\ce{CO2}}$, which is averaged both temporally and spatially in the lateral ($y$) direction. White lines denote positions where mean values of velocity and thermo-chemical quantities are extracted: (a) $x_q=-0.5H$, (b) $x_q=0$, (c) $x_q=0.5H$, and (d) $x_q=H$. Here, (b) $x_q=0$ corresponds to the streamwise position of the mean quenching point.}
\label{meanContour}
\end{figure}

\begin{figure}[h!]
\centering
\includegraphics[trim = 0mm 0mm 0mm 0mm, clip, width=240pt]{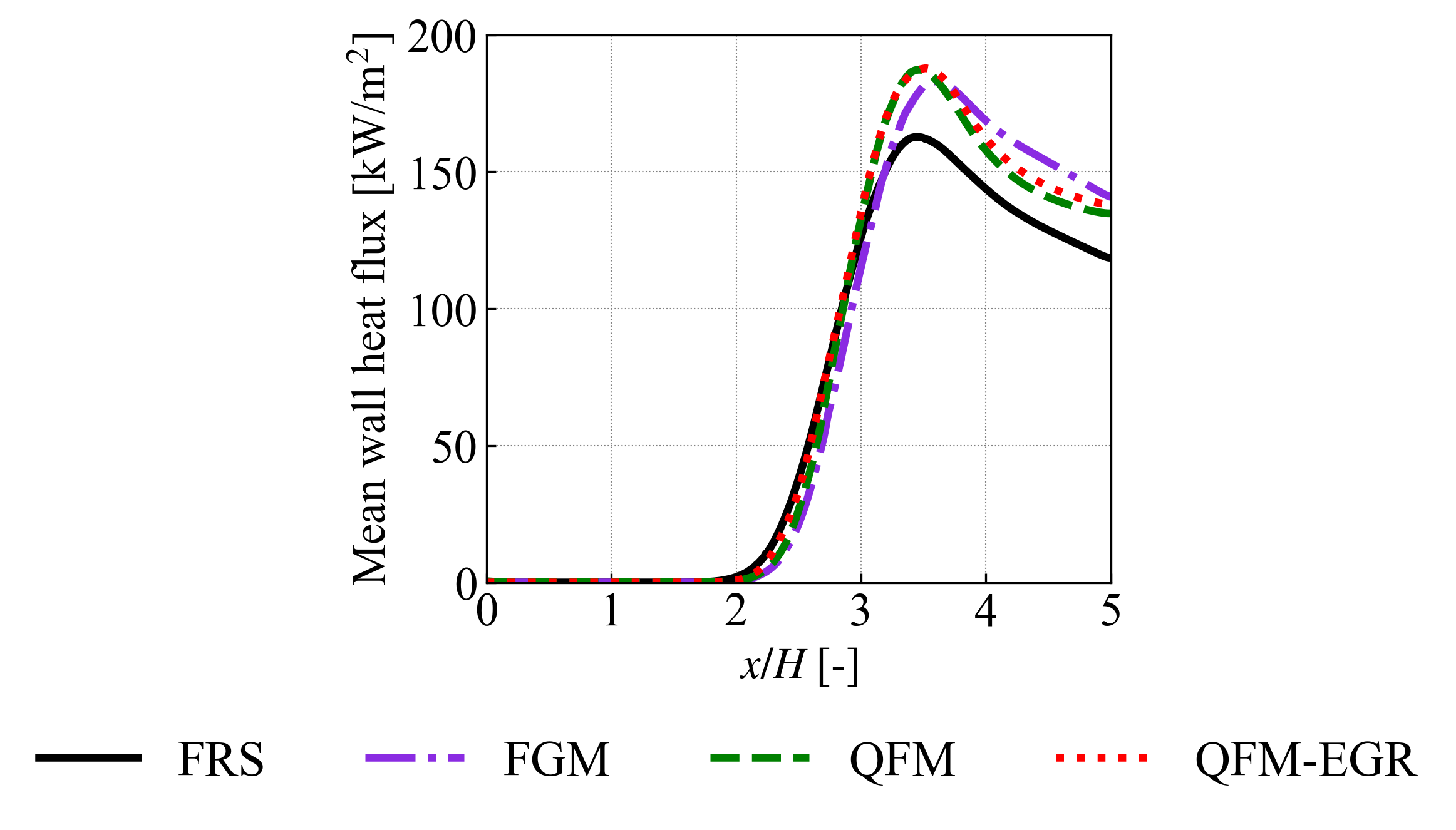}
\caption{Distributions of the mean wall heat flux along the normalized streamwise coordinate $x/H$. The reference FRS result is denoted by the black solid line. The violet, the green, and the red dashed lines correspond to FGM, QFM, and QFM-EGR results, respectively.}
\label{WallHeatFluxMean}
\end{figure}

The mean \ce{CO2} mass fraction is displayed in Fig.~\ref{meanContour}. It is observed that the overall flame appears similar comparing all the simulations. To enable a quantitative comparison between FRS, FGM, QFM, and QFM-EGR, the mean quenching points are first determined based on the profiles of the mean wall heat fluxes.
As shown in Fig.~\ref{WallHeatFluxMean}, the general trend observed for all simulations is that there is no heat transfer to the wall from the inlet until $x/H$ = 2, which is due to that the wall is assumed to be isothermal and its temperature is the same as the unburned gas. Afterwards, the mean wall heat flux gradually rises to reach a peak where the mean quenching point is defined, following which it decreases to a non-zero value until the outlet.
Overall, FGM, QFM, and QFM-EGR show greater heat transfer to the wall than the FRS. FGM, QFM, and QFM-EGR give similar predictions of the peak value, at 183.0 kW/m$^2$, 187.2 kW/m$^2$, and 187.7 kW/m$^2$, respectively. However, they are all higher than the reference value 162.7 kW/m$^2$, and the overprediction remains in the downstream region. This may be attributed to the deviations in $\lambda$ and the temperature gradient in the $z$ direction, according to the definition of the wall heat flux. Regarding the position where the wall heat flux reaches a maximum, it is quite close when comparing different simulations, at $x/H$ = 3.45, 3.6, 3.4625, 3.4625 for FRS, FGM, QFM, and QFM-EGR, respectively. The small differences are mainly originated from the quenching area, since the performances of FGM, QFM, and QFM-EGR are similar for the region away from the wall, see Figs.~\ref{UMean},~\ref{GlobalQuantities}, and \ref{speciesMean} below.

For a more detailed investigation, mean quantities of the velocity, temperature, enthalpy and progress variables are analyzed for four representative streamwise positions (white lines (a)-(d) in Fig.~\ref{meanContour}): 
\begin{itemize}
    \item (a) upstream of the quenching point ($x_q=-0.5H$)
    \item (b) at the quenching point ($x_q=0$)
    \item (c) downstream of the quenching point ($x_q=0.5H$)
    \item (d) close to the outlet ($x_q=H$).
\end{itemize}

\begin{figure}[h!]
\centering
\includegraphics[trim = 0mm 0mm 0mm 0mm, clip, width=470pt]{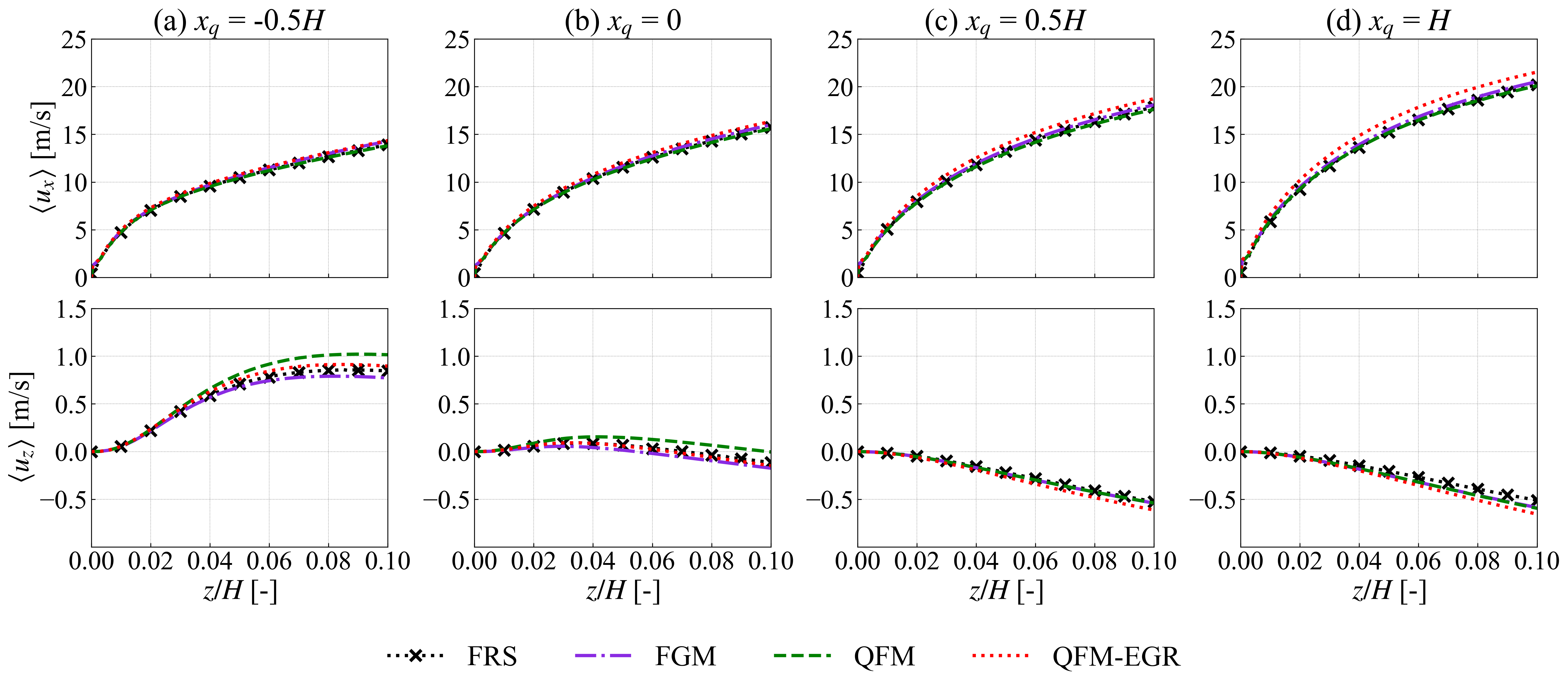}
\caption{Distributions of the mean streamwise velocity (top) and wall-normal velocity (bottom) from the bottom wall to the height of 0.1$H$ at different streamwise positions: (a) $x_q=-0.5H$, (b) $x_q=0$, (c) $x_q=0.5H$, and (d) $x_q=H$. The reference FRS results are denoted by black dashed lines marked with cross symbols. The violet, the green, and the red dashed lines correspond to FGM, QFM, and QFM-EGR results, respectively.}
\label{UMean}
\end{figure}

Figure~\ref{UMean} shows the streamwise velocity $u_x$ and the wall-normal velocity $u_z$ plotted along the wall-normal lines. For each position, the streamwise velocity is zero at the wall, which fulfills the no-slip boundary condition, and increases monotonically with increasing wall distance. The gradient of the streamwise velocity reaches a maximum at the wall, and gradually decreases along the wall-normal direction. Different trends are observed for the gradient of the wall-normal velocity, which remains at a low level close to the wall. In addition, the variation in the wall-normal velocity is different for the four positions considered. For position (a), the wall-normal velocity is positive and increases with the wall distance. For position (b), the wall-normal velocity remains almost zero from the wall until $z/H$ = 0.1. In contrast, a mainly negative wall-normal velocity is observed for positions (c) and (d), which is due to the thermal expansion of the burned gases, and the magnitude increases when moving away from the wall. This phenomenon is also observed in~\cite{heinrich_3d_2018}. These major characteristics are well captured by all manifolds. The absolute value of the wall-normal velocity is generally much smaller than that of the streamwise velocity, and the wall-normal velocity shows slightly larger discrepancies than the streamwise velocity at positions (a) and (b). However, overall, for both the streamwise and the wall-normal velocity, FGM, QFM, and QFM-EGR show good agreement with the reference data for all four streamwise positions considered.

\begin{figure}[ht!]
\centering
\includegraphics[trim = 0mm 0mm 0mm 0mm, clip, width=470pt]{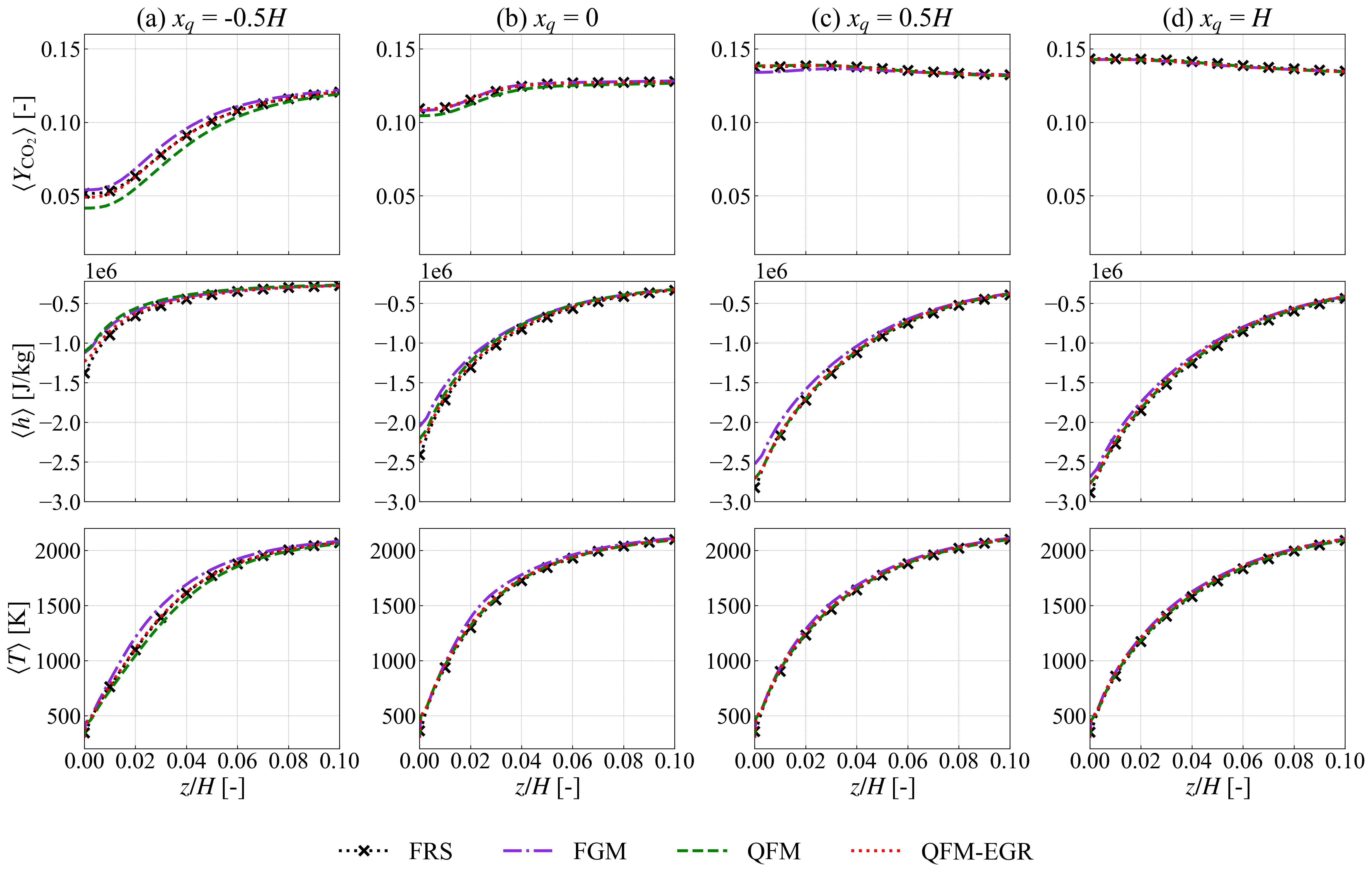}
\caption{Distributions of mean progress variable $Y_{\ce{CO2}}$ (top), enthalpy $h$ (middle), and temperature $T$ (bottom) from the bottom wall to the height of 0.1$H$ at different streamwise positions: (a) $x_q=-0.5H$, (b) $x_q=0$, (c) $x_q=0.5H$, and (d) $x_q=H$. The reference FRS results are denoted by black dashed lines marked with cross symbols. The violet, the green, and the red dashed lines correspond to FGM, QFM, and QFM-EGR results, respectively.}
\label{GlobalQuantities}
\end{figure}

Figure~\ref{GlobalQuantities} presents the results for the first progress variable, $Y_{\ce{CO2}}$, enthalpy, and the temperature. All flamelet manifolds yield profiles for the progress variable and temperature that are comparable to the reference. Small discrepancies are found in enthalpy distributions. The results from QFM-EGR are closest to the FRS for all positions. Meanwhile, QFM shows a slight overprediction in the near-wall region at position (a), and FGM leads to overprediction for all positions considered. 
The differences in the enthalpy profiles can be explained by the model limitations of the manifolds employed. The enthalpy boundary condition at the wall is set based on the species composition and the temperature, i.e., $h = h(Y_1,Y_2,...,Y_k,T,p)$~\cite{ketelheun2013heat}, and the species composition is retrieved from the tabulated manifolds. Therefore, any deviations in the tabulated thermo-chemical states will result in differences in the enthalpy boundary condition that are then reflected in the near-wall enthalpy profiles at the wall. Note that, however, the enthalpy set at the wall needs to be consistent with the manifold-based model employed to correctly capture the wall temperature.
Nevertheless, in summary, the results from the three flamelet manifolds do not show much difference.

\begin{figure}[h!]
\centering
\includegraphics[trim = 0mm 0mm 0mm 0mm, clip, width=470pt]{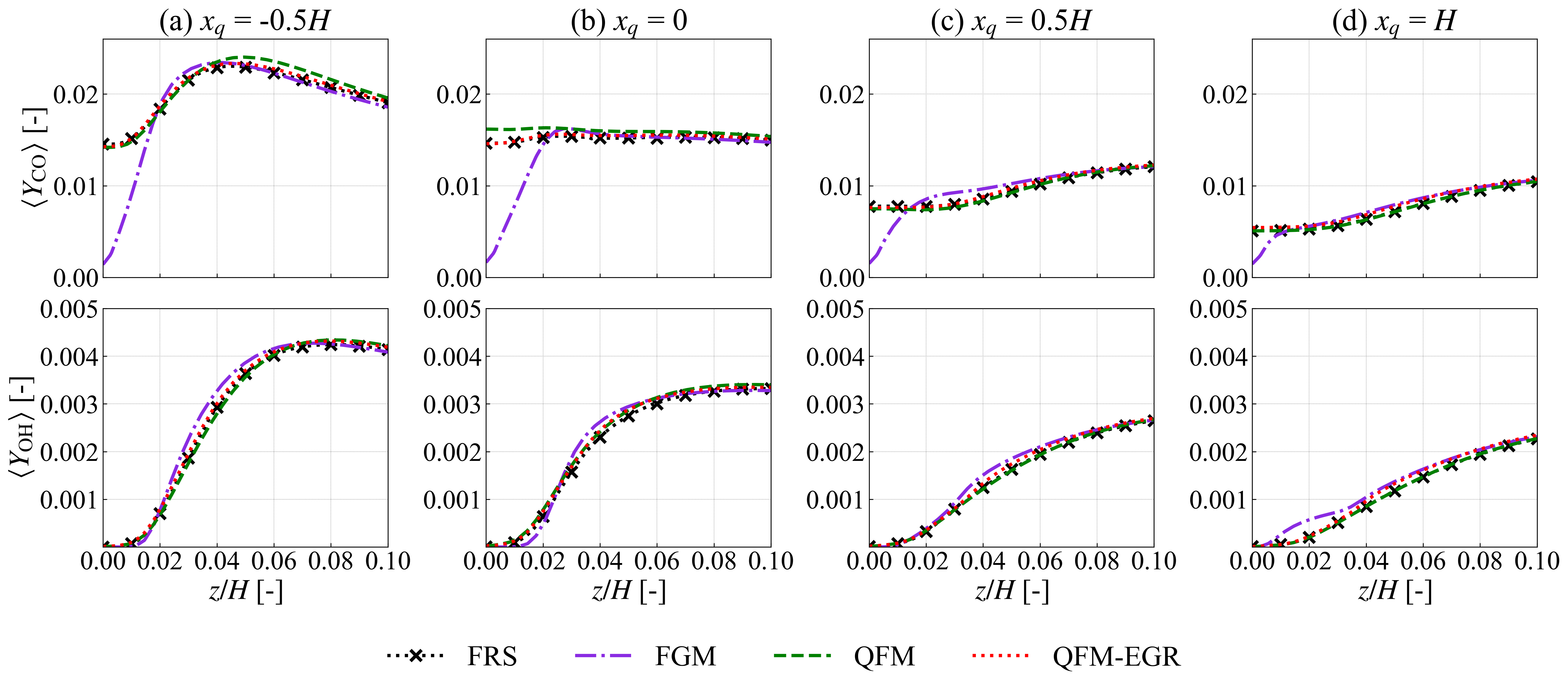}
\caption{Distributions of mean mass fractions of \ce{CO} (top) and \ce{OH} (bottom) from the bottom wall to the height of 0.1$H$ at different streamwise positions: (a) $x_q=-0.5H$, (b) $x_q=0$, (c) $x_q=0.5H$, and (d) $x_q=H$. The reference FRS results are denoted by black dashed lines marked with cross symbols. The violet, the green, and the red dashed lines correspond to FGM, QFM, and QFM-EGR results, respectively.}
\label{speciesMean}
\end{figure}

To assess the capability of the flamelet manifolds to predict the pollutants and radicals, distributions of mean \ce{CO} and \ce{OH} for the same four positions shown in Fig.~\ref{meanContour} are plotted in Fig.~\ref{speciesMean}. Accurate predictions can still be observed for QFM and QFM-EGR, with QFM-EGR performing slightly better, which is similar to the findings for major species. In contrast, deviations from the reference results are shown for FGM, regarding both \ce{CO} and \ce{OH}. In the case of \ce{CO} especially, even different variation trends are presented. For FRS, QFM, and QFM-EGR, the \ce{CO} distribution does not change much near the wall, while FGM presents a notable decrease towards the wall. Consequently, the near-wall \ce{CO} is significantly underpredicted by FGM, especially at positions (a) and (b), meaning that FGM is unable to reproduce the \ce{CO} accumulation here. Even at position (d), which is close to the outlet, FGM still underestimates the \ce{CO} at the wall. The deficiency of the FGM in capturing the near-wall \ce{CO} is consistent with previous laminar studies~\cite{ganter_laminar_2018, efimov_qfm_2020}, and it is originated from the improper \ce{CO} diffusion in the enthalpy direction included in the FGM table.

\begin{figure}[h!]
\centering
\includegraphics[trim = 0mm 0mm 0mm 0mm, clip, width=470pt]{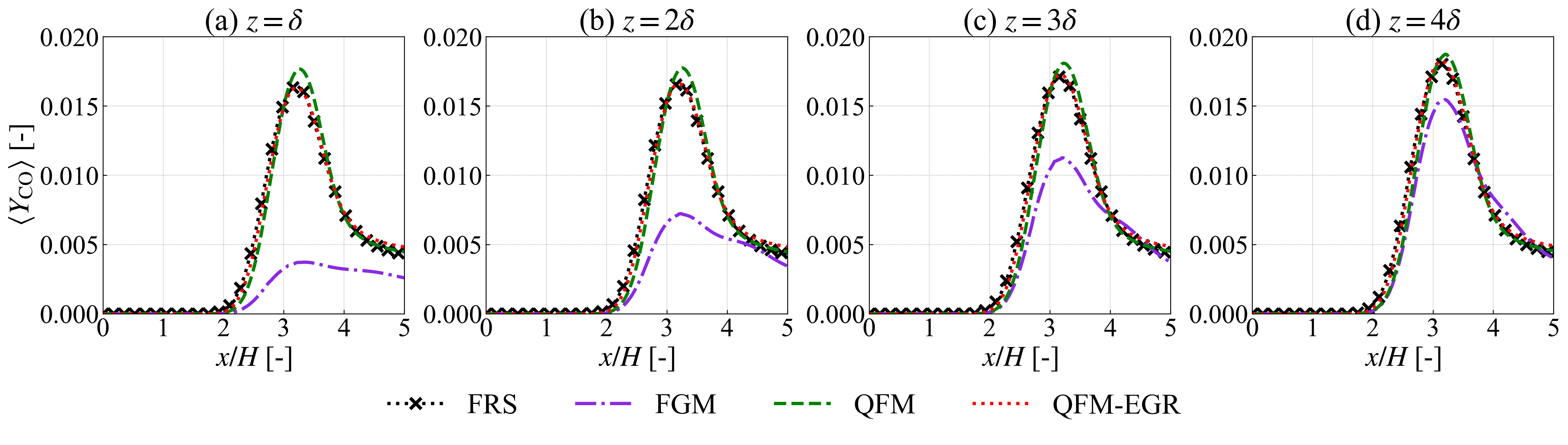}
\caption{Distributions of the mean \ce{CO} mass fraction from the inlet to the outlet at different vertical positions: (a) $z = \delta$, (b) $z = 2 \delta$, (c) $z = 3 \delta$, and (d) $z = 4 \delta$. The reference FRS results are denoted by black dashed lines marked with cross symbols. The violet, the green, and the red dashed lines correspond to FGM, QFM, and QFM-EGR results, respectively.}
\label{COMeanz}
\end{figure}
 
Further, the \ce{CO} mass fraction profiles in the streamwise direction for different wall distances are plotted in Fig.~\ref{COMeanz} to gain deeper insights into the \ce{CO} prediction. The minimum wall distance considered in Fig.~\ref{COMeanz} corresponds to the unstretched laminar flame thickness $\delta$. It is calculated using $\delta = \lambda_u/\left(\rho_u c_{p,u} s_L\right)$, where $\lambda_u$, $\rho_u$, and $c_{p,u}$ are the thermal conductivity, density, and heat capacity of the unburned gas, respectively, and $s_L$ denotes the laminar burning velocity. \ce{CO} is always underpredicted by FGM in the vicinity of the wall, e.g., position (a), which is consistent with previous studies~\cite{ganter_numerical_2017}. With increasing wall distance, the deviations in \ce{CO} prediction can still be found before the outlet, while the computed values at the outlet are close to reference results, such as positions (b), (c), and (d). Compared with FGM, results from QFM show evident improvement. This is similar to the findings in laminar SWQ cases \cite{efimov_qfm_2020,steinhausen2021numerical}. However, slight overprediction exists near the peak value. Compared to QFM, QFM-EGR provides an even better prediction, which shows excellent agreement with the FRS reference.

Based on the above discussions, it can be concluded that all three flamelet manifolds considered are capable of capturing the major global characteristics of the turbulent FWI. However, FGM shows significant deficiencies in the prediction of pollutants and radicals. These findings apply to both turbulent and laminar flames. Different from laminar cases, FVI exists in turbulent conditions. Therefore, the performance of the QFM and the recently introduced QFM-EGR specifically designed for turbulent FWI is further compared. It is observed that the latter shows a slight improvement in the prediction of the mean quantities.
In the following chapter, results from these two approaches will be comprehensively compared.

\subsection{Probability density functions of the CO mass fraction}  \addvspace{5pt}
\label{sec:PDF}

\begin{figure}[h!]
\centering
\includegraphics[trim = 0mm 0mm 0mm 0mm, clip, width=470pt]{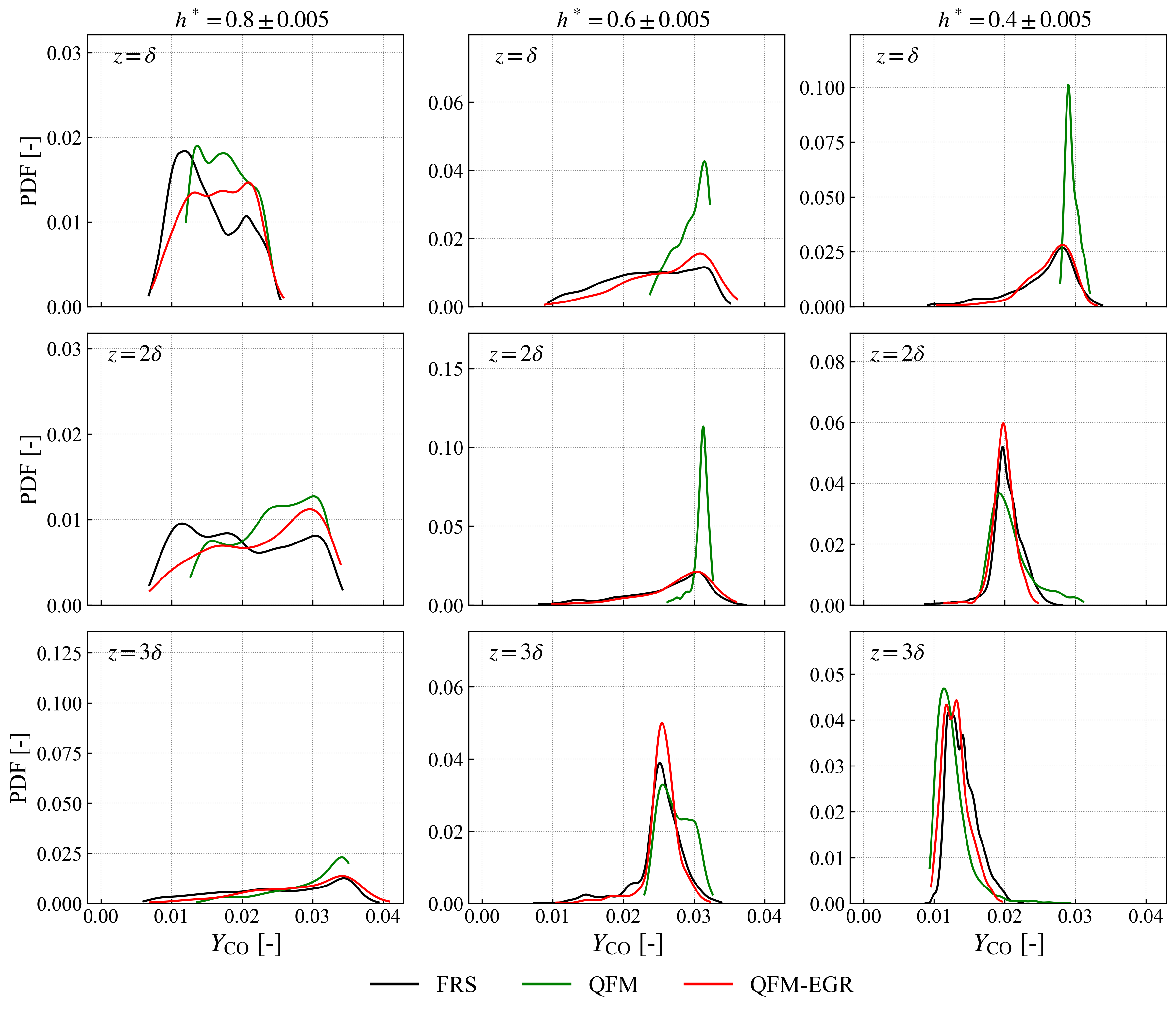}
\caption{PDFs of \ce{CO} conditioned on wall distance and enthalpy level.}
\label{COPDF}
\end{figure}

In addition to the mean quantities, the PDFs of the \ce{CO} mass fraction are analyzed in this section, as shown in Fig.~\ref{COPDF}. The PDFs are computed based on the resolved quantities and extracted at different wall distances ($z = \delta, \, 2\delta, \, 3\delta$) for a given enthalpy interval ($h^*$ = 0.8 $\pm$ 0.005, 0.6 $\pm$ 0.005, 0.4 $\pm$ 0.005). Here, $h^*$ denotes the normalized enthalpy $h^*=\left(h-h_\mathrm{min}\right)/\left(h_\mathrm{max}-h_\mathrm{min}\right)$, with $h^*$ = 1 corresponding to an undisturbed flame and $h^*$ = 0 to a fully quenched state. The data are sampled from 10 time instants.

Unlike the observations regarding the mean quantities, which reveal only small differences between QFM and QFM-EGR, significant and non-negligible differences are found in the PDFs of \ce{CO}, especially for positions very close to the wall. At the plane closest to the wall, i.e., $z=\delta$, the distribution of \ce{CO} predicted by QFM-EGR agrees quite well with the FRS.
The variation range of \ce{CO} is almost the same when comparing QFM-EGR and FRS. However, the PDF of QFM is much narrower, and concentrates at higher \ce{CO} values. This corresponds to the overprediction of the peak of the mean \ce{CO} shown in Fig.~\ref{COMeanz}. The deviation between QFM and QFM-EGR increases with increasing heat loss. For low enthalpy levels such as $h^*$ = 0.6 and $h^*$ = 0.4, a distinct peak can be observed in the PDF of QFM, while the PDF distribution is much wider in QFM-EGR and FRS. Moving further away from the wall to $z=2\delta$, the discrepancies between QFM and QFM-EGR still exist, especially for $h^*$ = 0.6. However, for high enthalpy losses, e.g., $h^*$ = 0.4, the PDF of CO narrows, leading to a better agreement between QFM and QFM-EGR. This means that the enthalpy range where QFM and QFM-EGR differ significantly becomes smaller with increasing wall distance.
At $z=3\delta$, the PDFs from QFM and QFM-EGR are quite similar. 
Therefore, it is reasonable to deduce that there would be almost no differences between QFM and QFM-EGR in the case of positions far away from the wall. 
Based on the above findings, it can be concluded that the significant differences in the PDFs of \ce{CO} between QFM and QFM-EGR mainly exist in the region near the flame tip, which is located close to the wall at an intermediate level of enthalpy losses. According to \cite{zentgraf_classification_2021,steinhausen2022flame}, this is also the area where the FVI mechanism plays a role. Therefore, it is assumed that the characteristics of the PDFs shown above are related to this mechanism. To verify this, the FVI mechanism will be studied in the following sections.

\subsection{Prediction of flame-vortex interaction}  \addvspace{5pt}
\label{sec:influencesOfFVI}

Firstly, the instantaneous thermo-chemical states that are directly influenced by the FVI mechanism are compared for FRS, QFM, and QFM-EGR, so that the importance of the mechanism can be illustrated. Afterwards, the flame dynamics of an FVI event using QFM-EGR is compared to the FRS based on a time-series analysis to further evaluate the performance of QFM-EGR in capturing the FVI mechanism.

\subsubsection{Comparison of thermo-chemical states}
\label{sec:thermoChemicalStates}

To investigate the instantaneous local thermo-chemical states, quantities are evaluated over 10 time steps. For each time instant, data are collected from 31 independent slices parallel to the $xz$ plane. For each slice, the region of interest covers a range of ($x_Q$ - 50$\delta$, $x_Q$ + 50$\delta$) in $x$ direction and (0, $z_Q$ + 10$\delta$) in $z$ direction, as denoted by the green rectangle in Fig.~\ref{configuration}.

\begin{figure}[h!]
\centering
\includegraphics[trim = 0mm 0mm 0mm 0mm, clip, width=470pt]{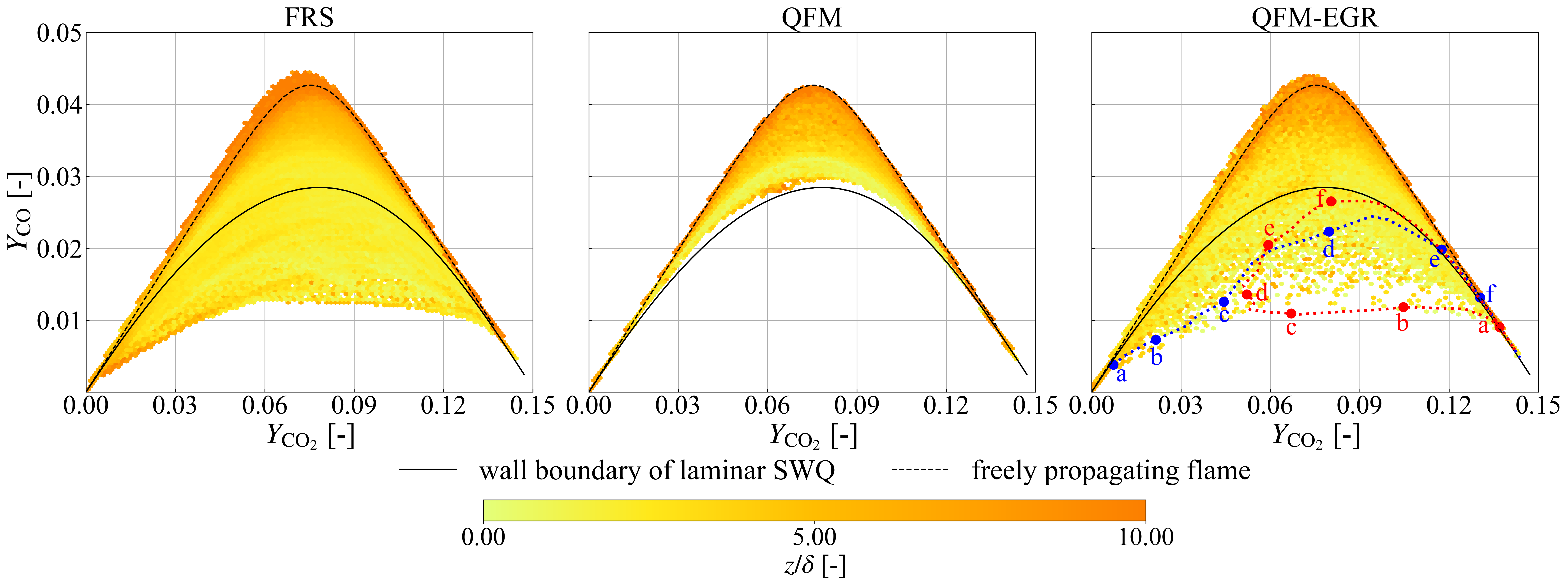}
\caption{Thermo-chemical states near the quenching point for different lateral positions and time instants: (left) FRS, (middle) QFM, and (right) QFM-EGR. The scatters are colored by the normalized wall distance $z/\delta$. For reference, the states of a 1D freely propagating flame and on the wall boundary of a laminar SWQ are denoted by black dashed lines and black solid lines, respectively.}
\label{scatterTotal}
\end{figure}

\begin{figure}[h]
\centering
\includegraphics[trim = 3mm 2mm 2mm 2mm, clip, width=450pt]{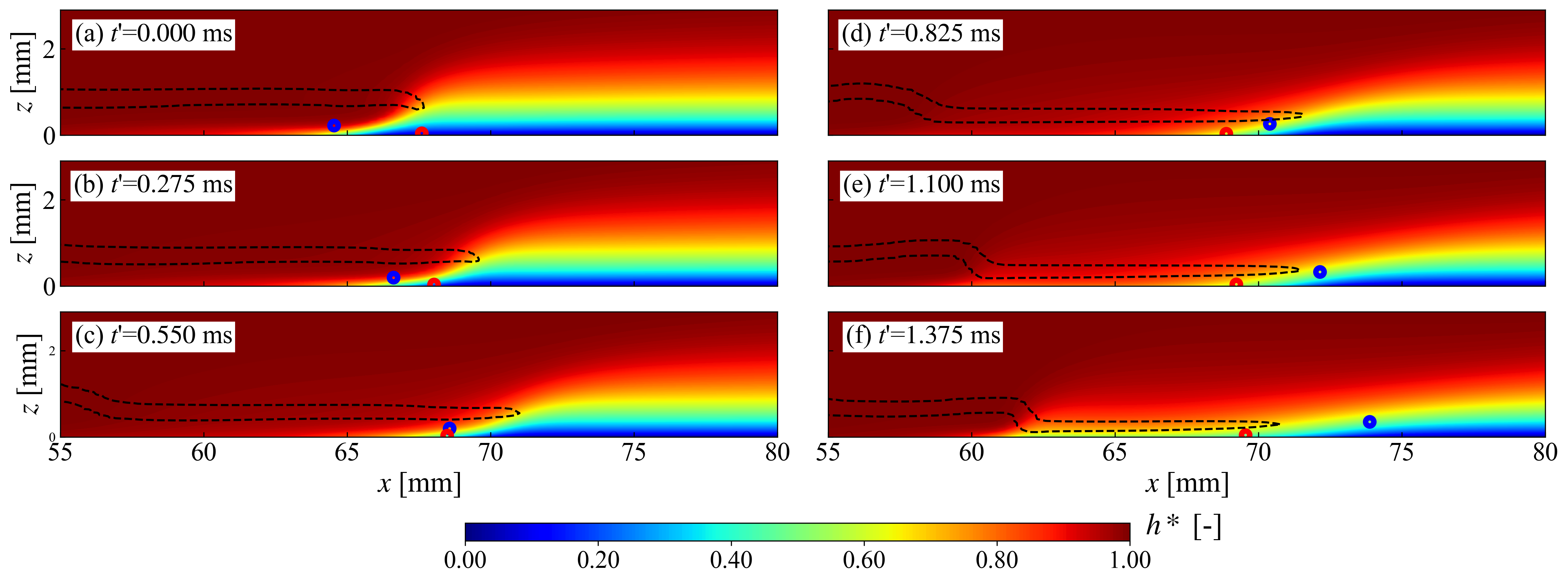}
\caption{Time series of a slice in the lateral direction through the turbulent flame for QFM-EGR. Contours of the normalized enthalpy ($h^*$) are shown. Isocontours of heat release rate (0.1 $\times$ HRR$_\mathrm{max}$) are denoted by black dashed lines. The blue and red points correspond to the positions of the Lagrangian particles projected to the $xz$ plane where the two particles initially locate.}
\label{hContour}
\end{figure}

Figure~\ref{scatterTotal} shows the thermo-chemical states in \ce{CO2}-\ce{CO} space for the region of interest at different lateral positions and times instants, with results from FRS on the left, QFM in the middle, and QFM-EGR on the right. The data are colored by the normalized wall distance ($z/\delta$). Additionally, the thermo-chemical states on the wall boundary of a corresponding laminar SWQ are shown as black solid lines. Here, the laminar SWQ simulation is performed on a two-dimensional domain, similarly to \cite{steinhausen2021numerical}.
The thermo-chemical states of a freely propagating flame are denoted by black dashed lines, where the composition and the temperature of the fresh gas remain the same as the current turbulent SWQ configuration.
From the FRS results, it is observed that the state space spanned by the turbulent SWQ can not be fully covered by its laminar counterpart, as shown by the area surrounded by the black solid and dashed lines in Fig.~\ref{scatterTotal}. Especially for the region very close to the wall, i.e., the low-temperature region, the minimum value of the conditional \ce{CO} mass fraction is significantly below the limit of the wall boundary of the laminar SWQ, namely $\mathrm{min}\left(Y_{\ce{CO}} | Y_{\ce{CO2}}\right) < \left(Y_{\ce{CO}} | Y_{\ce{CO2}} \right)_{\mathrm{laminar, z=0}}$. According to \cite{zentgraf_classification_2021,steinhausen2022flame}, this is due to cooled burned products mixing with fresh gases in the vicinity of the quenching point.
In the study by Steinhausen et al.~\cite{steinhausen2022flame}, the QFM-EGR was introduced to incorporate the mixing effect in the thermo-chemical state and validated a priori. As a result, the LES with the QFM-EGR is able to correctly capture the local mixing processes caused by FVI in the turbulent flame, showing a large portion of the scatters below the limit of the wall boundary of the laminar SWQ. 
To further illustrate the mixing process, the evidence of FVI in the physical space is additionally shown for better understanding. Here, Lagrangian massless particles are placed in the flow field. Note that more details about the flow field, e.g., Q-criteria, can be found in the supplementary material and also previous studies~\cite{steinhausen2022flame, zentgraf_classification_2021}.As an example, the trajectories of two Lagrangian particles in \ce{CO2}-\ce{CO} space are additionally highlighted in the scatters of QFM-EGR in Fig.~\ref{scatterTotal}. Initially, the two particles locate on the same $xz$ plane. Their corresponding trajectories in the physical space projected onto the initial $xz$ plane for 6 representative time instants are indicated in Fig.~\ref{hContour}, where a slice of the flame in the lateral direction is shown for QFM-EGR. Note that the movement in the lateral direction is minor during these 1.375~ms in the case of both particles. A relative time $t'$ is introduced and $t'=0$ refers to the first time instant considered. Initially, the blue particle is located in the unburned part of the flame, while the red one is in the burned region. In the following, the particles move closer to each other until they almost collide at $t'$ = 0.55 ms, consistent with the mixing process proposed in~\cite{zentgraf_classification_2021, steinhausen2022flame}. Note that the movement of the red particle is against the main flow direction due to the interaction with the turbulent vortices, which can be observed from its location relative to the flame tip. This is also reflected in the thermo-chemical states shown in Fig.~\ref{scatterTotal}, where both particles exhibit CO values far below the laminar limit that is also the lower bound of the QFM accessible range. After $t'$ = 0.825 ms, the thermo-chemical states of both particles move in the direction of the burned gas. 
Based on the above observations, it can be concluded that FVI leads to a mixing process of burned and fresh gases in the near-wall region, which results in thermo-chemical states that are not present in the laminar cases. This has been found both in simulations~\cite{steinhausen2022flame} and experiments~\cite{zentgraf_classification_2021}. Therefore, the capability of QFM-EGR to correctly predict FVI in the turbulent FWI is demonstrated. The QFM, on the other hand, only accounts for the strong heat losses due to the flame quenching at the wall. This is the reason for the much narrower state space spanned by the LES with QFM and the narrower PDF distributions in Fig.~\ref{COPDF}, since the states originating by the turbulent mixing are not included in the manifold. This means that QFM can not capture the mixing process in the near-wall region, because there is no variation in the composition of the unburned gas in the flamelets used for the manifold generation. This model limitation is overcome by the QFM-EGR. 
Considering the notable difference between QFM and QFM-EGR in Fig.~\ref{scatterTotal}, the importance of the FVI mechanism is verified. Therefore, in the prediction of transient thermo-chemical states in the near-wall region for turbulent conditions, the benefit of the QFM-EGR becomes evident.

\subsubsection{Evolution of the FVI area}
\label{sec:timeSeries}

\begin{figure}[h]
\centering
\includegraphics[trim = 3mm 2mm 2mm 2mm, clip, width=450pt]{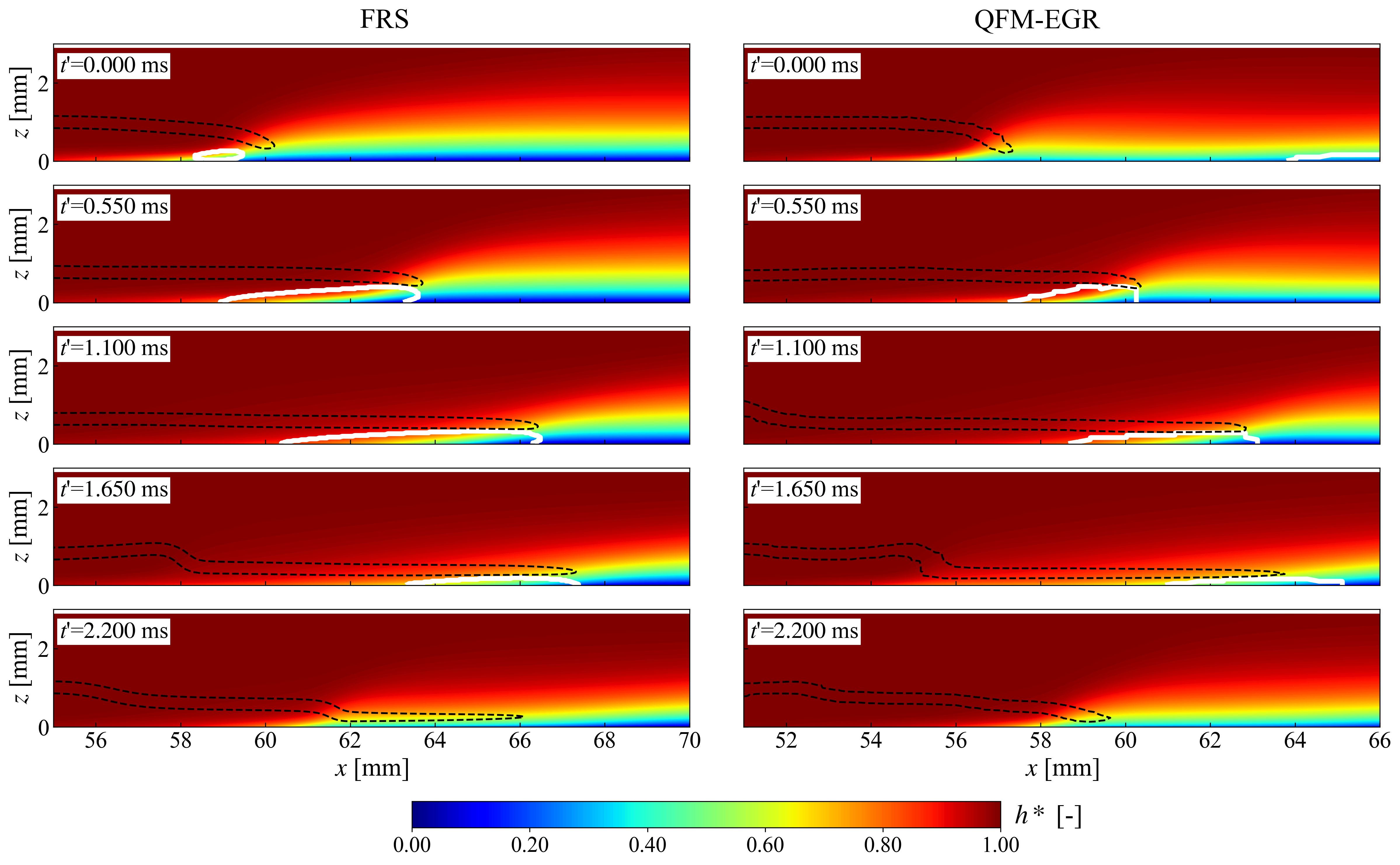}
\caption{Time series of a slice in the lateral direction through the turbulent flame: (left) FRS, (right) QFM-EGR. Contours of the normalized enthalpy ($h^*$) are shown. Isocontours of heat release rate (0.1 $\times$ HRR$_\mathrm{max}$) are denoted by black dashed lines. The white isocontour represents the area of FVI.}
\label{CO2Contour}
\end{figure}

To further demonstrate the capability of QFM-EGR to correctly capture the FVI mechanism, the time evolution of the FVI area is investigated in this section. A time series for a slice of the flame in the lateral direction for both FRS and QFM-EGR are depicted in Fig.~\ref{CO2Contour}. Here, isolines of the heat release rate (0.1 $\times$ HRR$_\mathrm{max}$) are denoted by black dashed lines. The white isolines corresponds to $\left(Y_{\ce{CO}} | Y_{\ce{CO2}}\right) = \left(Y_{\ce{CO}} | Y_{\ce{CO2}} \right)_{\mathrm{laminar, z=0}}$. The states of the region surrounded by the white isolines fulfill the condition that $\left(Y_{\ce{CO}} | Y_{\ce{CO2}}\right) < \left(Y_{\ce{CO}} | Y_{\ce{CO2}} \right)_{\mathrm{laminar, z=0}}$. According to Section~4.3.1 and~\cite{zentgraf_classification_2021,steinhausen2022flame}, this is caused by the FVI. Therefore, this region is called FVI region, which is related to the turbulent flow structures.
The time series from the FRS is chosen to display the typical behavior found in the turbulent SWQ flame, similarly to~\cite{steinhausen2022flame}.
For QFM-EGR, results from different time instants are used, so that they present similar flame movements to FRS. For each simulation, the time interval of the neighboring two slices stays at 0.55 ms. Similarly to the last section, the relative time $t'$ is also introduced here; $t'=0$ refers to the first time instant considered for each simulation in Fig.~\ref{CO2Contour}. Following~\cite{heinrich_large_2018}, all flame angles less than 2$^{\circ}$ are categorized as HOQ-like events, while other cases are classified as SWQ-like scenarios.
Based on FRS results, it is observed that the flame is in an SWQ-like state at $t'=0$, and FVI takes place in only a very small area. Afterwards, the angle between the flame and wall decreases and the FVI area begins to grow, e.g., at $t'$ = 0.55 ms and 1.1 ms. As the flame-wall impact angle decreases further, the FVI region begins to shrink. Consequently, an HOQ-like event occurs at $t'$ = 1.65 ms, with the FVI mechanism only playing a role in a very small region close to the wall. In the following, the quenching scenario transfers to SWQ-like again, as can be observed for $t'$ = 2.2 ms. At this time instant, no FVI region is detected. A repeated flame behavior between the HOQ-like scenario and the SWQ-like scenario is observed, so the evolution after $t'$ = 2.2~ms is not shown for brevity. Readers interested in additional time instants may refer to the supplementary materials of ~\cite{steinhausen2022flame}. For all time instants when FVI occurs, it is observed that the FVI area decreases with increasing wall distance, both in physical space and enthalpy space. This helps to explain the findings from Fig.~\ref{COPDF} that significant differences between QFM and QFM-EGR are mainly seen in the vicinity of the wall, and the enthalpy range involved becomes smaller when moving further away from the wall. These major characteristics of the flame dynamics can also be found in QFM-EGR, as shown in the right-hand column of Fig.~\ref{CO2Contour}. According to \cite{steinhausen2022flame}, the repeated flame behavior is caused by the interaction between the flame and the vortex. Therefore, it indicates that QFM-EGR is able to correctly capture the dynamic evolution of the FVI mechanism.

\section{Conclusion} \addvspace{10pt}
\label{sec:Conclusions}
In the current study, three flamelet manifolds with different levels of complexity are applied to the LES of a turbulent SWQ flame with the purpose of comprehensively evaluating their predictive capability in \textit{a-posteriori} calculations. These help to provide guidance for practical applications. In FGM, heat losses are considered without the inclusion of species gradients in the enthalpy direction. This drawback is remedied by QFM, where heat losses to the wall are taken into account based on HOQ flamelets. QFM-EGR combines QFM with EGR to further consider the flame dilution, following our previous work~\cite{steinhausen2022flame}. Models are assessed with the results from an FRS as a reference, and the following conclusions can be drawn:
\begin{itemize}
    \item Regarding the mean quantities, all three flamelet manifolds are good at predicting general flame characteristics, such as the mean flow and temperature fields, as well as major species. However, the flamelet manifolds perform differently in the prediction of pollutants and radicals. FGM presents significant deviations from the reference results in the near-wall region. In contrast, results from QFM and QFM-EGR show great improvement, with QFM-EGR performing slightly better.
    \item In the case of the PDFs of \ce{CO}, significant differences are observed between QFM and QFM-EGR. Compared to QFM, the distribution of \ce{CO} covers a wider range in QFM-EGR, which shows better agreement with the reference. 
    \item Looking into the local thermo-chemical states, scatters of the QFM results are found to be restricted within the laminar counterpart, while QFM-EGR results span a much wider space, including a large portion of scatters below the limit of the wall boundary of the laminar SWQ. The results from QFM-EGR are more consistent with the FRS, indicating the importance of the FVI mechanism in the transient near-wall behavior. Moreover, the FVI mechanism is also illustrated in physical space with injected Lagrangian massless particles.
    \item To further verify the capability of QFM-EGR to capture the FVI mechanism, the time evolution of the FVI is investigated. It is found that a flame dynamic similar to that shown in FRS also exists in QFM-EGR.
\end{itemize}

In conclusion, the FGM considering varying enthalpy levels, is the simplest manifold to build and accurately predicts general flame quantities, such as the flow field, temperature, and major species. However, it fails in the prediction of pollutants and radicals. For this purpose, the QFM shows an improved prediction accuracy at the cost of more complex manifold generation, while the manifold dimensions remain unchanged and the computational cost thus stays at a similar level. The QFM-EGR shows the overall best prediction accuracy by capturing the influence of FVI at the cost of an additional table dimension and thus increased computational cost and memory requirements. When these manifolds are applied in real combustors, the benefits and drawbacks of each model should be considered when choosing the manifold.


\section*{Acknowledgments} \addvspace{10pt}
The authors thank Dr. Thorsten Zirwes from the group of Prof.~Henning Bockhorn (Karlsruhe Institute of Technology) for supplying the data on the flame-resolved simulation with detailed chemistry.  This research is supported by Deutsche Forschungsgemeinschaft (DFG, German Research Foundation) within SFB/Transregio 150 (project number 237267381), the Darmstadt Graduate School of Excellence Energy Science and Engineering (GSESE), and the European Union’s Horizon 2020 research and innovation programme under the Center of Excellence in Combustion project, grant agreement No 952181. Calculations for this research were conducted on the Lichtenberg high-performance computer at TU Darmstadt.

\bibliography{publication.bib}
\bibliographystyle{unsrtnat_mod}

\newpage
\section*{Supplementary material} \addvspace{10pt}

\begin{figure}[h]
\centering
\includegraphics[trim = 3mm 2mm 2mm 2mm, clip, width=470pt]{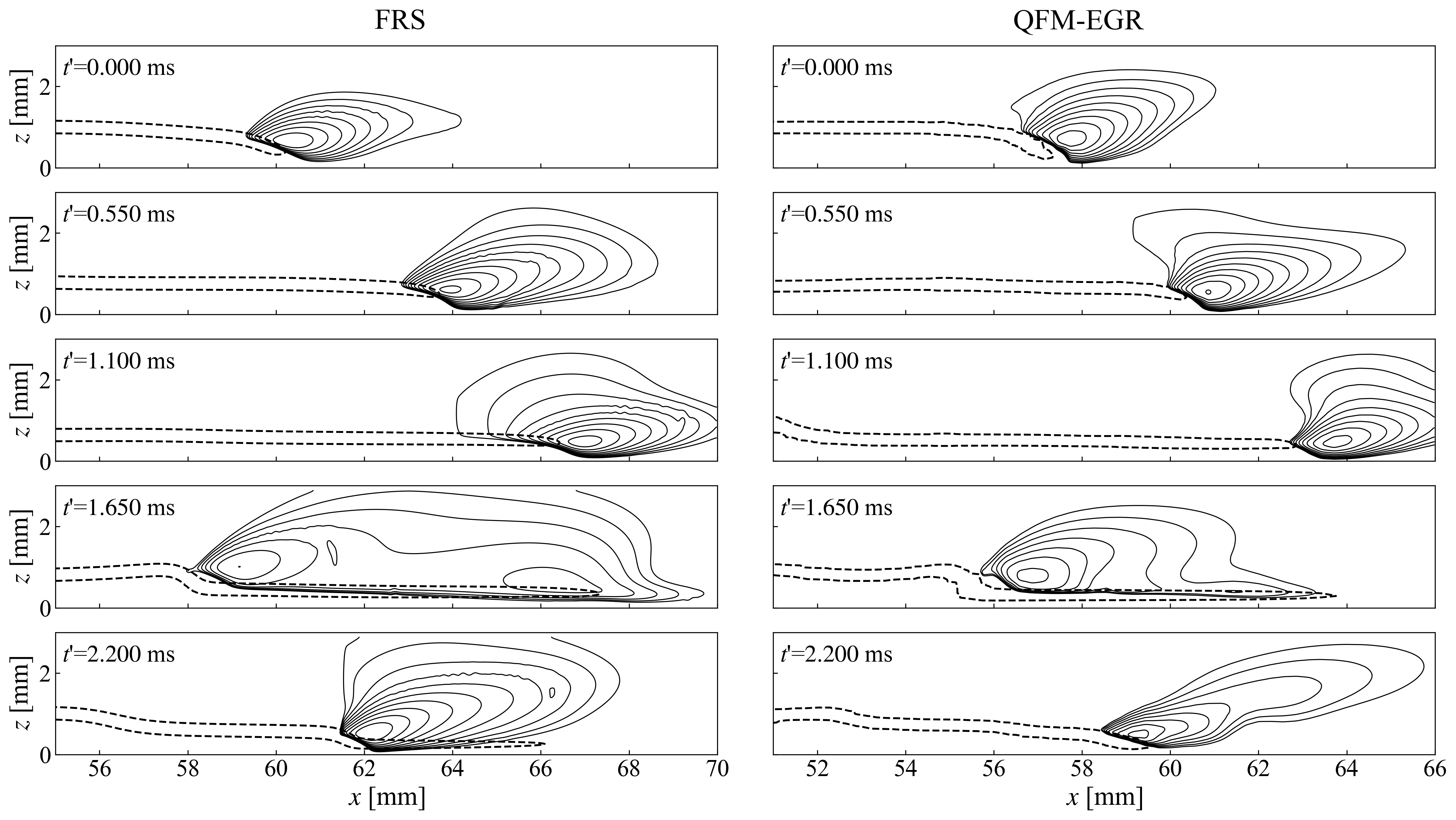}
\caption{Time series of a slice in the lateral direction through the turbulent flame: (left) FRS, (right) QFM-EGR, corresponding to Fig.~11 in the main text. Vortical structures are visualized by the Q-criterion (black solid lines). Isocontours of heat release rate (0.1 $\times$ HRR$_\mathrm{max}$) are denoted by black dashed lines.}
\label{QCriterion}
\end{figure}

\end{document}